\documentclass[manuscript,screen]{acmart}  
\usepackage{graphicx}
\usepackage{amsmath}  
\usepackage[linesnumbered,ruled,vlined]{algorithm2e}
\AtBeginDocument{%
  \providecommand\BibTeX{{%
    \normalfont B\kern-0.5em{\scshape i\kern-0.25em b}\kern-0.8em\TeX}}}

\setcopyright{acmcopyright}
\copyrightyear{2023}
\acmYear{2023}
\acmDOI{XXXXXXX.XXXXXXX}

\acmConference[fashionxrecsys’23]{17th ACM Conference on Recommender Systems, Singapore}{18th-22nd September 2023}{Singapore}

\begin{document}

\title[Diversify and Conquer]{Diversify and Conquer: Bandits and Diversity for an Enhanced E-commerce Homepage Experience}






\author{Sangeet Jaiswal}
\affiliation{%
  \institution{Myntra Designs Pvt Ltd}
  \city{Bangalore}
  \country{India} }
\email{sangeet.jaiswal@myntra.com}

\author{Korah T Malayil}
\affiliation{%
  \institution{Myntra Designs Pvt Ltd}
  \city{Bangalore}
  \country{India}
  }
\email{korah.malayil@myntra.com}

\author{Saif Jawaid}
\affiliation{%
  \institution{Myntra Designs Pvt Ltd}
  \city{Bangalore}
  \country{India}
  }
\email{saif.jawaid@myntra.com}

\author{Sreekanth Vempati}
\affiliation{%
  \institution{Myntra Designs Pvt Ltd}
  \city{Bangalore}
  \country{India}
  }
\email{sreekanth.vempati@myntra.com}

\renewcommand{\shortauthors}{Jaiswal et al.}

\begin{abstract}
In the realm of e-commerce, popular platforms utilize widgets to recommend advertisements and products to their users. However, the prevalence of mobile device usage on these platforms introduces a unique challenge due to the limited screen real estate available. Consequently, the positioning of relevant widgets becomes pivotal in capturing and maintaining customer engagement.
Given the restricted screen size of mobile devices, widgets placed at the top of the interface are more prominently displayed and thus attract greater user attention. Conversely, widgets positioned further down the page require users to scroll, resulting in reduced visibility and subsequent lower impression rates. Therefore it becomes imperative to place relevant widgets on top. However, selecting relevant widgets to display is a challenging task as the widgets can be heterogeneous, widgets can be introduced or removed at any given time from the platform. 
In this work, we model the vertical widget reordering as a contextual multi-arm bandit problem with delayed batch feedback. The objective is to rank the vertical widgets in a personalized manner. We present a two-stage ranking framework that combines contextual bandits with a diversity layer to improve the overall ranking. We demonstrate its effectiveness through offline and online A/B results, conducted on proprietary data from Myntra, a major fashion e-commerce platform in India. 

\end{abstract}





\maketitle

\section{INTRODUCTION}
 In the world of e-commerce, the website or app real estate holds significant value and serves as a key catalyst in shaping user experience, engagement, and conversion rates. Most of the homepages have distinct sections, comprising various widgets that feature diverse content such as product recommendations, monetization products and ads, promotions, and sale details. As the number of widgets increases, the demand for a systematic and relevant display order becomes more pronounced. Our data reveals that, on average, users scroll down the homepage up to three times and click on widgets within the first two scrolls. 
 
Addressing this challenge posed a few hurdles in our context, Widgets are owned by different teams, and slotting them properly requires domain knowledge and validation through A/B tests, which is time-consuming and also not feasible sometimes. The notion of a widget ID is not tangible since the same widget can be represented with different widget IDs in different layouts. Widgets are also heterogeneous in nature, for example, they can be banners, awareness campaigns, product carousels, engagements (videos), order tracking cards, etc. The contents of these widgets are powered by different recommendation models and are personalized to the user. Hence the representation of these widgets needs to be learned without taking the content into consideration. Specifically, the particular brands or products within these widgets are not available as signals to the ranker due to system constraints.

For this reason, we have engineered new features such as Core theme, Widget intent, etc. to identify what type of content is being showcased through the widgets. These have been instrumented as mandatory fields to be filled by business teams while creating/deploying new widgets and layouts.
Existing approaches to tackle ranking problems like this rely on having sufficient historical data\cite{5694074, DBLP:journals/corr/ChengKHSCAACCIA16} to predict user preferences. However, a limitation arises when there is minimal data about a new content item, hindering effective recommendations. This issue, known as the ``item cold-start problem'' arises especially when introducing novel item types to a platform.
To overcome the item cold-start problem and provide effective recommendations, the Multi-Armed Bandit (MAB) approach\cite{thompson1933likelihood} is proposed. MAB balances the exploitation of known user preferences with the exploration of new items. Contextual bandit algorithms, a type of MAB, utilize context information, including user and item representations, to achieve improved performance. By assuming a linear relationship between the expected reward and context, contextual bandit algorithms can enhance recommendations in cold-start scenarios with limited data about newly added items\cite{abbasi2011improved,chu2011contextual}.

\textbf{Contributions.} In this research paper, we utilize the contextual bandit framework to tackle the challenge of optimizing the order of widgets in order to enhance the customer experience on our mobile shopping app. Our contributions include:
\begin{itemize}

\item Implementing a contextual bandit framework with diversity for dynamic widget reordering in real-time. 
\item Demonstrate superiority of bandit learning in a dynamic environment by comparing the Linear UCB model against a more complex supervised XGBoost model.

\end{itemize}

\begin{figure}[htbp]
    \centering
    \includegraphics[scale=0.4]{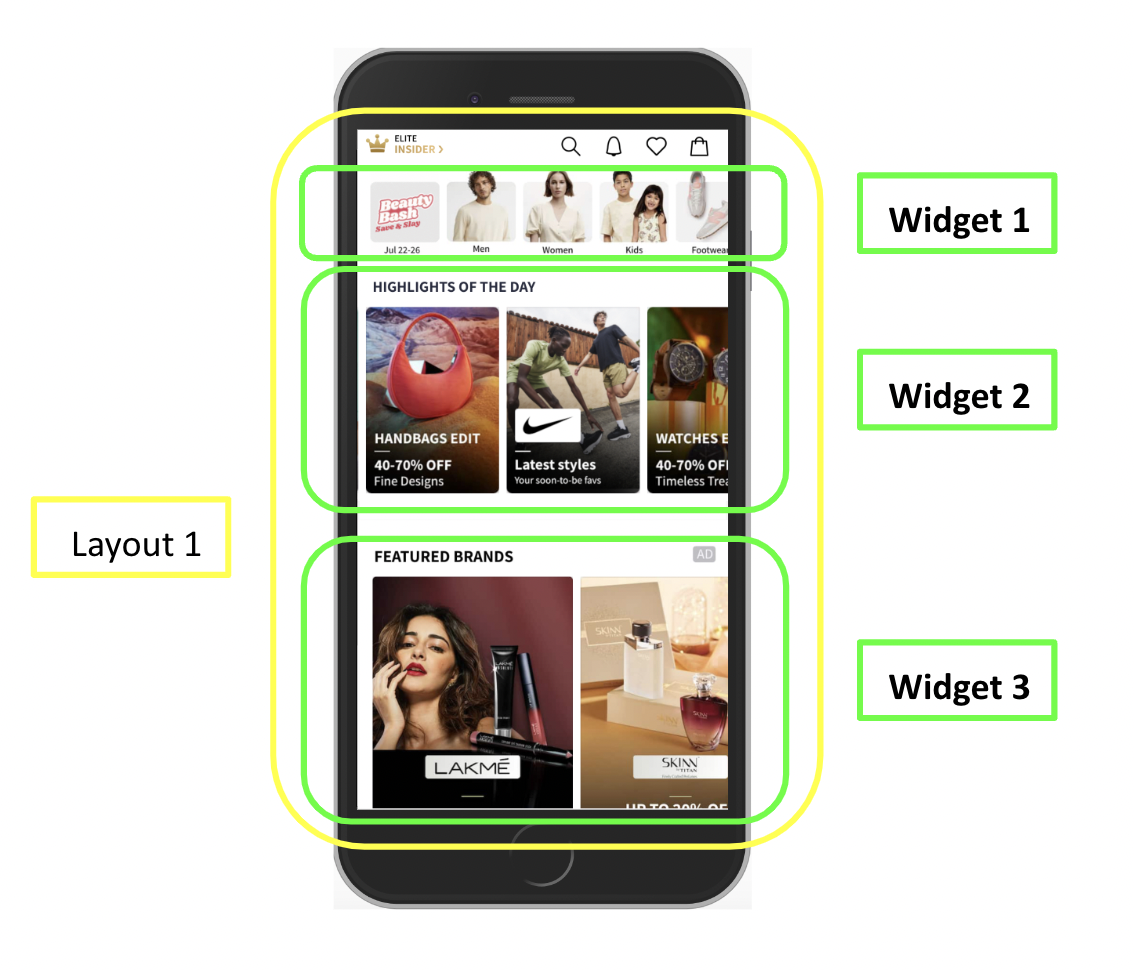}
    \caption{Example of widgets and layout on our home page: A different set of users can be assigned to different layouts, each with its own set of widgets that may not be mutually exclusive. }
    \label{fig:layout_widgets}
\end{figure}

\section{RELATED WORK}
Ever since multi-armed bandits\cite{thompson1933likelihood} were introduced, there has been substantial research and industry implementation in various domains such as personalized recommendations\cite{li2010contextual}, dynamic product pricing\cite{misra2019dynamic} and portfolio optimisation\cite{huo2017risk}. Multi-armed bandits are algorithms that make decisions over time by balancing the exploration and exploitation of different options(arms) such that the cumulative reward over time gets maximized. Contextual bandits are an extension of the multi-armed bandit problem \cite{pandey2007bandits}\cite{NIPS2007_4b04a686}\cite{wang2005bandit} where at each round the decision maker has access to some context related to the arms before deciding which arm to chose. These have emerged as a powerful approach in recent years, outperforming traditional recommender systems by effectively addressing the challenge of cold start cases. One notable algorithm in this context is LinUCB, which was introduced by \cite{li2010contextual}. It assumes a linear payoff model, where the expected payoff or reward of an action is a linear function of the associated context features. LinUCB has also shown good empirical performance in \cite{DBLP:journals/corr/KrishnamurthyAD15} even when the problem was not linear.
To obtain rankings from traditional MAB, we make the assumption that each widget is regarded as an independent arm and use its score for ranking. However, this assumption might not hold in practice, as widgets could have interdependencies.
Another approach could involve creating individual Multi-Armed Bandits (MABs) for each ranking position. In this method, we would treat each widget as an arm within the MAB. The MABs' outcomes would be influenced by the previously selected arms, as displaying the same widget multiple times would not be allowed, similar to the principles of the Ranked Bandit Algorithm (RBA) \cite{radlinski2008learning}. Nevertheless, this approach also falls short of adequately exploring the interdependencies among widgets and requires managing multiple algorithms simultaneously. In our approach, we have represented each widget with an embedding following the suggestion mentioned in \cite{saito2022off} to capture interdependencies among widgets. Recently, there have been notable studies exploring various approaches to address ranking problems, Kangas et al. investigated the use of Evolutionary algorithms to model combinatorial bandits to tackle such problems\cite{kangas2022scalable}. In another research, the integration of Deep Neural Networks with LinUCB was explored to get a better representation of context features \cite{10.1145/3543873.3587684}. However, they did not delve into the analysis of diversity in their results. 






\section{METHODOLOGY}
Let $W$ be the set of all possible candidate widgets, and let $l$ be the total number of slots to be shown on the page. Here, the cardinality of $W$ $\gg$ $l$. The content inside each widget is owned by different teams and ranking within these widgets is not controlled by this algorithm. Two prominent challenges that we faced in ranking widgets are,
First, widgets are dynamic in nature. Widgets are constantly created and removed on the platform in an ongoing manner. We also observed that customer preferences and their interaction with widgets change over time. Conventional recommendation systems trained continuously tend to recommend widgets that closely resemble those with which the user has previously engaged. This leads to no exploration and only exploitation using users' preferences. To mitigate these challenges, we formulate the problem of ranking widgets as a Multi-arm Bandit (MAB) problem.  

\begin{table}[htbp]
    \centering
        \begin{tabular}{|c|p{4cm}|c|p{4cm}|}
        \hline
        \textbf{Notation} & \textbf{Explanation} & \textbf{Notation} & \textbf{Explanation}   \\
        \hline
        $W$ & Set of all possible widgets & $l$ & Total number of slots \\
        $\textbf{u}$ & User Context  & $\textbf{w}$ & Widget Context \\
        $S$ & item-item similarity matrix & $C$ & Set of all customers \\
        $\mathbb{D}_t$ & Batch feedback & $\alpha$ & Exploration parameter \\
        $\gamma$ & forgetting factor in LinUCB & $\theta$ & Kernel parameter of DPP to control the relevance and diversity \\

        \hline
        \end{tabular}
        \caption{Notation used in this paper.}
        \label{tab:notation_table}
\end{table}

\subsection{Contextual Bandit Framework}
Specifically, within MAB, we use contextual bandits, where context contains both the user's and the widget's features. The User context, denoted as $\textbf{u}$, comprises a $d$-dimensional user vector $u_v \epsilon R^d$ which captures the user preference on the platform, including user browsing, add to card (ATC), and purchase history. Different time ranges are used while generating embeddings to capture users' short-term and long-term preferences. To construct $\textbf{u}$ we draw inspiration from prod2vec\cite{vasile2016meta} which employs a similar approach to generate product representation. Additionally, we leverage a $k$-dimensional user representation $u_{svd} \epsilon R^k$ learned through the Singular Value Decomposition (SVD) decomposition of the matrix constructed using user and widget interactions. 
Similarly, Widget context denoted as $\textbf{w}$, encompasses a m-dimensional widget vector $w_v \epsilon R^m$. This vector is created using features derived from widget meta-information and widget attributes. Furthermore, we leverage SVD decomposition to learn a $k$-dimensional representation of each widget $w_{svd} \epsilon R^k$. Apart from user and widget features, we have also taken other contextual features such as `hour of the day', `day of the week'(dow), `day of the year'(doy), and so on. User revenue-related features are also incorporated into the model for example `avg no of sessions per week', `avg bucket size' etc.

We combine $\textbf{u}$, $\textbf{w}$, and other features to form a single $z$-dimensional context vector $\textbf{x} \epsilon R^z$.

\begin{figure}[htbp]
    \centering
    \includegraphics[scale=0.5]{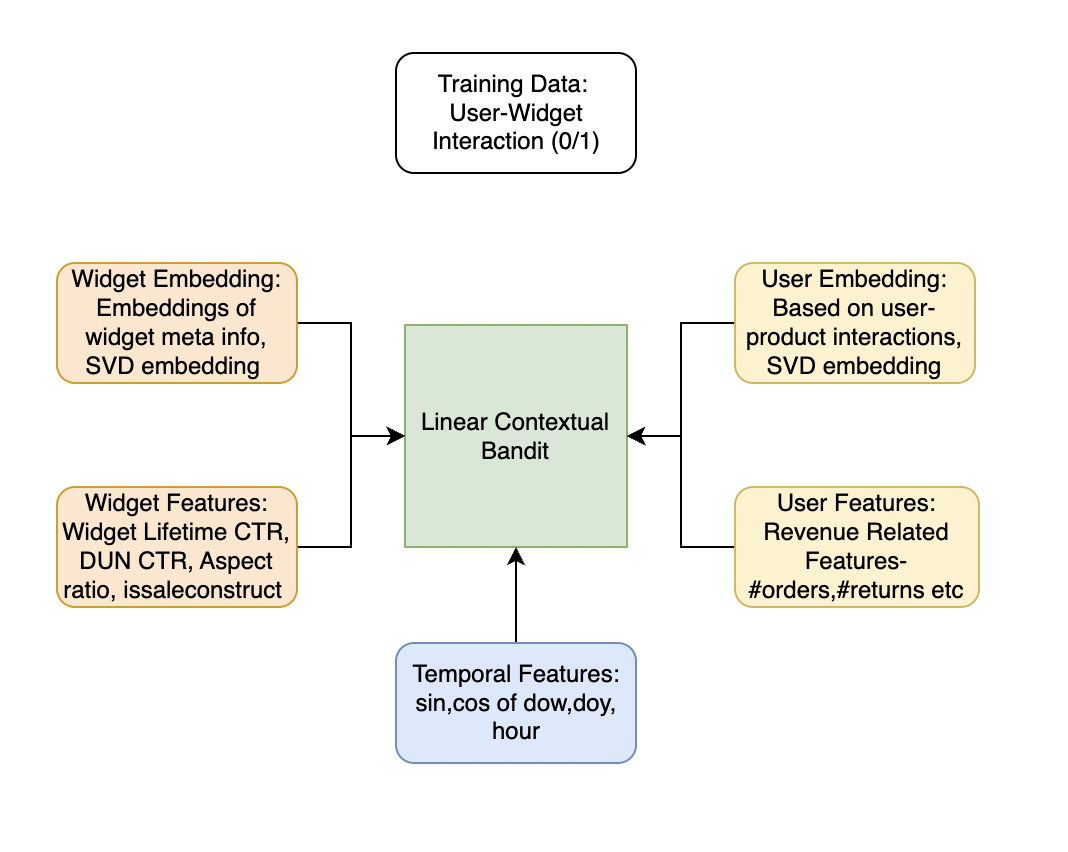}
    \caption{Our Bandit Model in detail}
    \label{fig:data_usage_diagram}
\end{figure}


In this paper, we have explored Linear contextual bandits particularly the LinUCB and Linear Thompson Sampling (LinTS)\cite{pmlr-v28-agrawal13}. 
In practice, due to resource and run-time constraints, user-widget interactions are not processed immediately, and model parameters are updated using batch feedback. This means that instead of updating the model after each individual interaction, we collect interactions over a certain period, and then update the model parameters using this aggregated feedback.
Both LinUCB and LinTS possess similar representation capabilities. While LinUCB deterministically selects arms until new feedback is incorporated, LinTS adopts a stochastic approach by sampling from a posterior distribution, allowing it to randomize over actions even in the absence of updated rewards\cite{bendada2020carousel}. This difference in exploration-exploitation strategies between LinUCB and LinTS forms the basis for their comparison in this study. 
Now once we have the context vector in place, CB works as follows:
\begin{itemize}

\item In any round t, let $K_t$ be the number of widgets in play and we have contextual features for each widget  $x_{t, i}$.
\item The algorithm makes a recommendation of the current round t, based on the parameters learned using contexts $x_i$ and rewards $r_i$ for actions $a_i$ $ i \epsilon [1,2,...,t-1]$ in previous rounds and the current context $x_t$.
\item The algorithm receives the rewards of action $a_t$, which are based on clicks and impressions.

\end{itemize}

In our specific setting, where we represent each widget with embedding, widgets that share similar attributes might possess highly similar representations. As a result, the bandit algorithm could end up selecting similar widgets, leading to reduced diversity in the chosen widgets. This issue arises due to the linear bandit's focus on optimizing the expected reward without explicitly considering diversity as an optimization criterion. Consequently, the algorithm may tend to exploit a narrow subset of widgets that have demonstrated favorable performance and widgets having similar representations, limiting its exploration of diverse alternatives. In Table \ref{tab:my_table} we have demonstrated the impact of exploration coefficient $\alpha$ on diversity metrics, as detailed in section $4.2.1$. The table showcases the variations in these metrics corresponding to different values of $\alpha$. To compute these metrics we have taken the top 10 recommendations generated by the model. 
Although $\alpha$ plays a crucial role in the exploration-exploitation trade-off, our results indicate that diversity metrics exhibit relative stability across different values of this coefficient. 

Note that the reward estimation function is the same for all the arms. This allows us to effectively handle the dynamic number of arms.

\begin{table}[h]
    \centering
    \begin{tabular}{|c|c|c|}
        \hline
        Model & ILAD & ILMD \\
        \hline
        LinUCB ( $\alpha=0.01$ ) & 0.48024 & 0.04021 \\
        LinUCB ( $\alpha=0.05$ ) & 0.48103 & 0.03926 \\
        LinUCB ( $\alpha=0.25$ ) & 0.48079 & 0.03942 \\
        LinTS ( $\alpha=0.01$ ) & 0.48118 & 0.03959 \\
        LinTS ( $\alpha=0.05$ ) & 0.48733 & 0.04127 \\
        LinTS ( $\alpha=0.25$ ) & 0.50219 & 0.04185 \\
        \hline
    \end{tabular}
    \caption{Impact of exploration coefficient $\alpha$ on diversity metrics}
    \label{tab:my_table}
\end{table}

\subsection{Incorporating Diversity using DPP}
Focusing solely on high-relevance items might result in a repetitive selection of similar products, disregarding the diverse preferences and interests of individual users. The lack of content diversity can hinder the exploration of potential appealing options, thus diminishing the overall engagement and satisfaction of customers during their shopping journey.
In this paper, we proposed a two-stage ranking model for widgets ranking. In the first stage, we leverage a Contextual Bandit model to generate relevance scores for the widgets. Subsequently, we employ the Determinantal Point Process (DPP) to re-rank the widgets based on their relevance scores and widget features. DPP balances the relevance of widgets and their similarities. Our approach is outlined in the Algorithm \ref{algo:linucb}, which is used to rank widgets. The observed rewards corresponding to the actions are logged and used to retrain the model using batch feedback. To account for changes in the environment\cite{graepel2010web} and to give more importance to recent data samples, we incorporate a forgetting factor $\gamma \epsilon [0,1] $ during model parameter update. Algorithm \ref{algo:dpp} is adopted from \cite{chen2018fast} which presents a fast implementation of the greedy MAP inference algorithm for DPP where they have defined diversity in the feature space of the entire subset unlike techniques based on pairwise dissimilarities. We followed a similar approach as highlighted in the paper to create the kernel matrix using $p_{t,1:K_t}$ and $X_{t,1:K_t}$. DPP selects a subset of items based on stopping criteria so $Y_t \subseteq K_t$. For the final ranking, we append the remaining widgets based on the scores obtained from the CB algorithm.


    


\begin{algorithm}
\SetAlgoLined
\caption{LinUCB with diversity using DPP}
\label{algo:linucb}
\KwIn{Number of Rounds $T$, Number of widgets eligible per round $K_t$, Dimension of feature vectors $z$, exploration parameter $\alpha$}
\KwOut{Recommended ordered widgets}
\BlankLine
\textbf{Initialize:} Set $\alpha > 0$, $A \leftarrow I_z$, $b \leftarrow 0_z$;
\BlankLine
\For{$t = 1, 2, \ldots, T$}{
    $\phi_t \leftarrow A^{-1}_t b_t$\;
    $\mathbb{D}_t=\emptyset$;
    
    Given a user we observe $K_t$ features, $x_{t,1},x_{t,2},...x_{t,K_t} \epsilon \mathbb{R}^z$
    
    \For{$k = 1$ \KwTo $K_t$}{
        Calculate the upper confidence bound (UCB): $p_{t,k} \leftarrow \phi_t^T x_{t,k} + \alpha \sqrt{x_{t,k}^T A_t^{-1} x_{t,k}}$\;
    }
    Ranked widgets $Y_t \leftarrow DPP(p_{t,1:K_t}, X_{t,1:K_t})$
    
    recommend widgets $Y_t$, observe rewards $R_t$\;
    $\mathbb{D}_t=\mathbb{D}_t \cup (X_t,Y_t,R_t)$

    Update the model using batch feedback $\mathbb{D}_t$
}
\end{algorithm}

\begin{algorithm}
\SetAlgoLined
\caption{DPP}
\label{algo:dpp}
\KwIn{Positive semi-definite kernel matrix $L \in \mathbb{R}^{K_t \times K_t}$, $stopping$ $criteria$ }
\KwOut{A subset of items $Y_t$}
\BlankLine
\textbf{Initialize:}  $c_i = []$, $d^2_i=\boldsymbol{L}_{i,i}$, $j=\text{argmax}_{i \in K_t} \log(d^2_i)$, Set $Y_t = \{j\}$\;
\BlankLine
\While {stopping criteria not satisfied}{
\For{$i \in K_t \setminus Y_t$}{
   $e_i = (\textbf{L}_{j,i} - (\textbf{c}_j \cdot \textbf{c}_i))/d_j$\;
   
   $\textbf{c}_i$ = [$\textbf{c}_i$, $e_i$]; $d^2_i = d^2_i - e^2_i$\;
}
$j = \text{argmax}_{i \in K_t \setminus Y_t} \log(d^2_i)$\;

$Y_t = Y_t \cup \{j\}$\;
}

\end{algorithm}




\begin{figure}[h]
    \centering
        \begin{minipage}[t]{0.45\linewidth}
            \centering
            \includegraphics[scale=0.2]{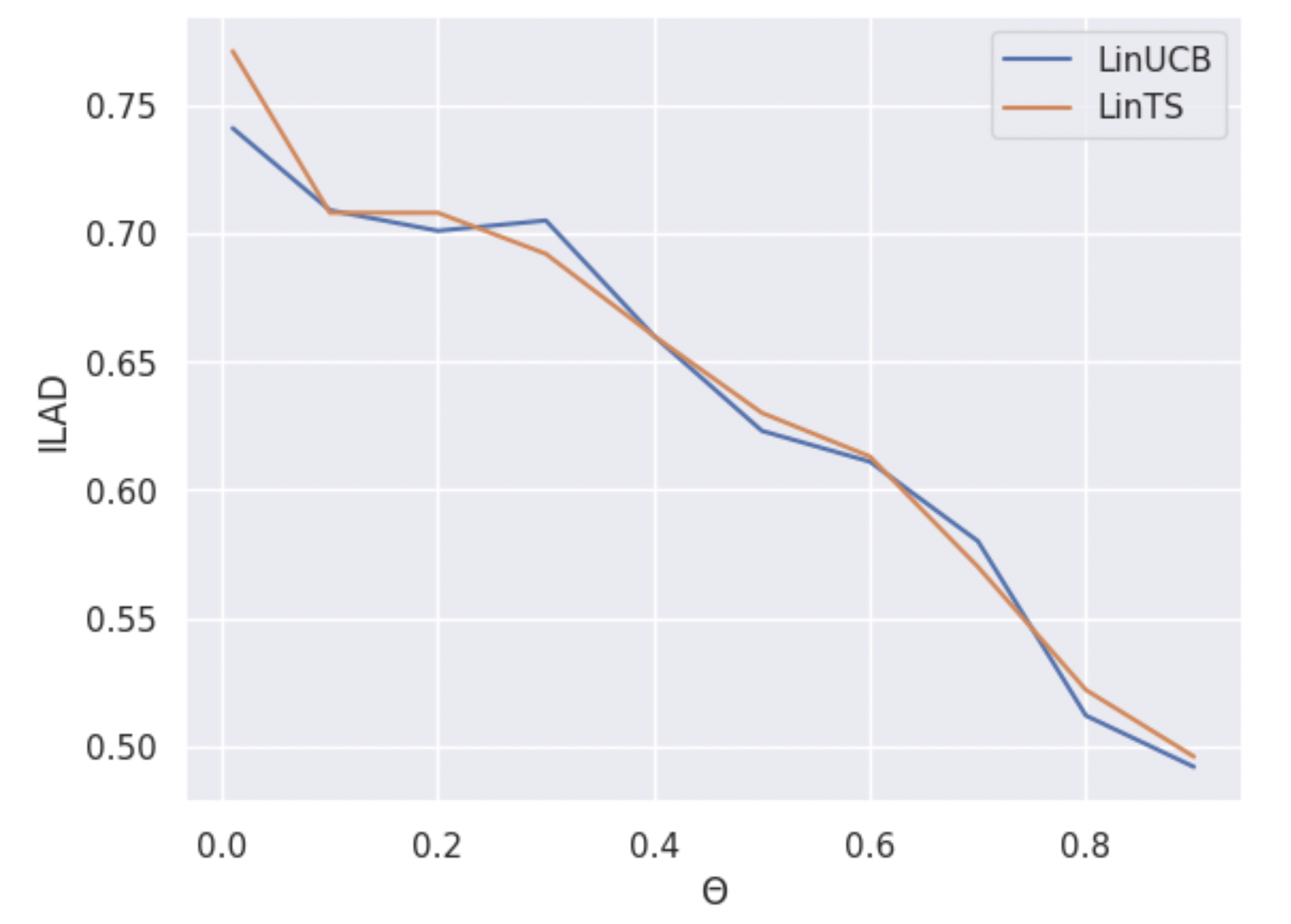}
            \label{fig:ILAD}
        \end{minipage}
        \hfill
        \begin{minipage}[t]{0.45\linewidth}
            \centering
            \includegraphics[scale=0.2]{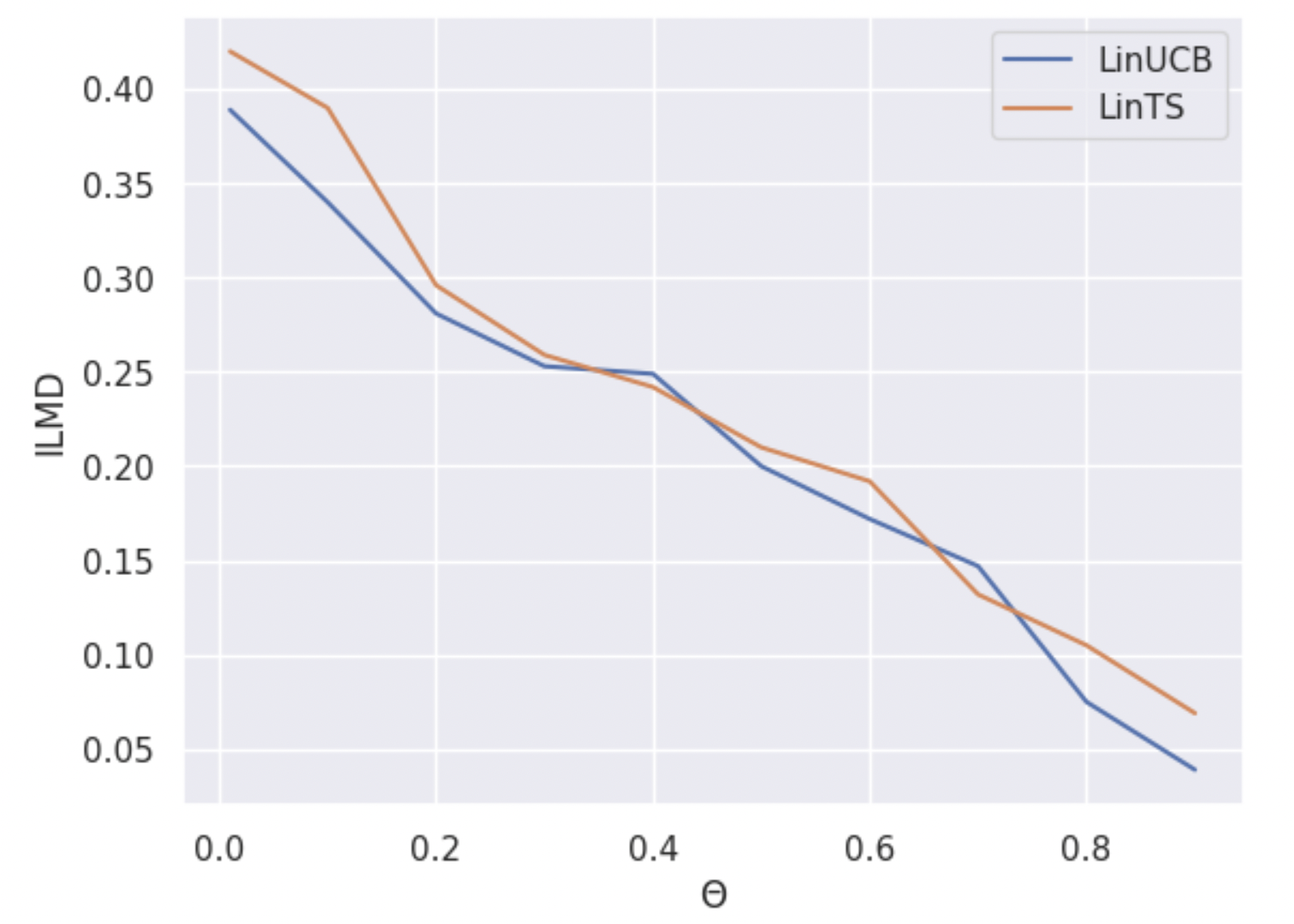}
            \label{fig:ILMD}
        \end{minipage}
        
        \begin{minipage}[t]{0.45\linewidth}
            \centering
            \includegraphics[scale=0.2]{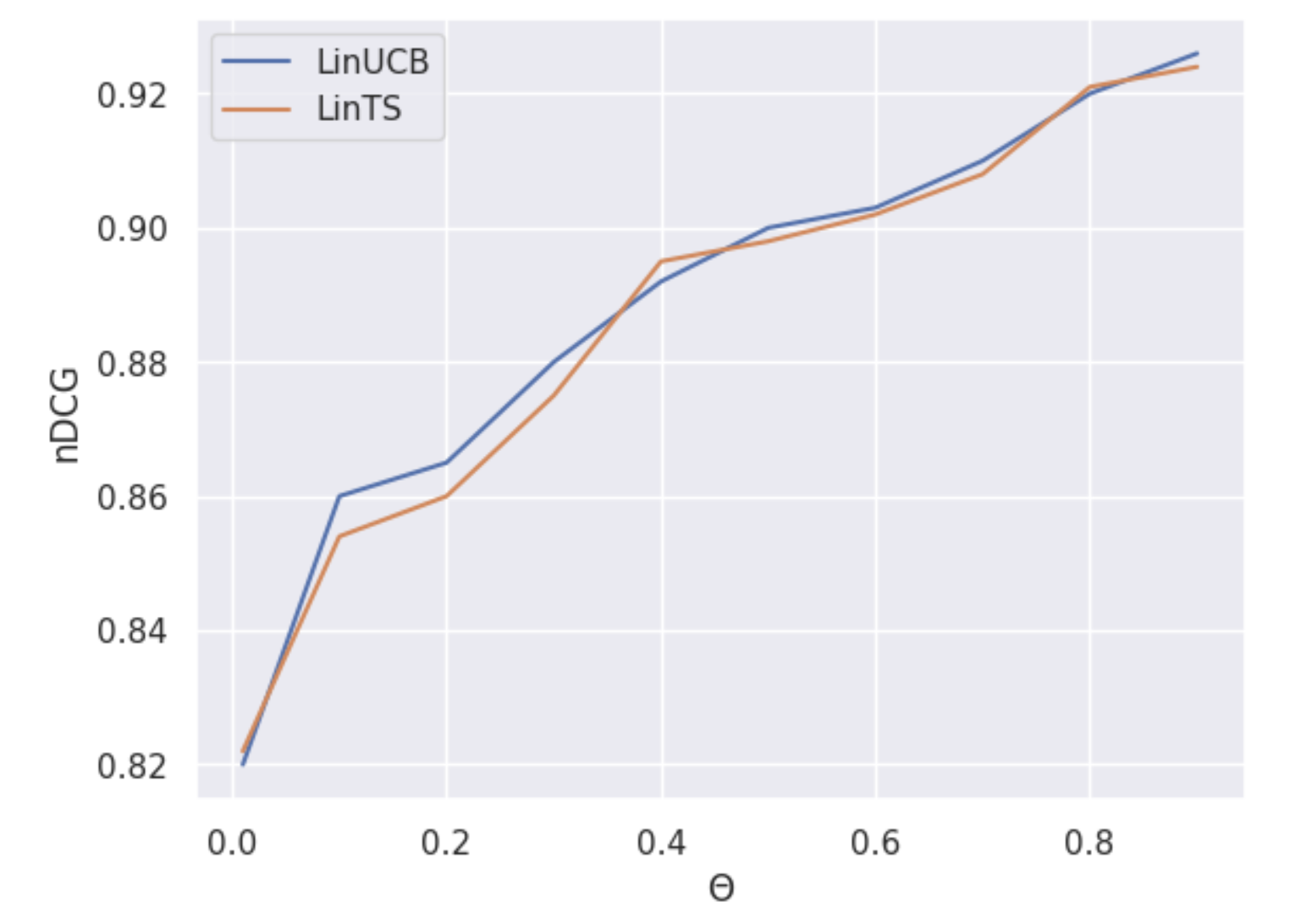}
            \label{fig:nDCG}
        \end{minipage}

    \caption{Influence of $\theta$ on relevance (nDCG) and diversity metrics (ILAD, ILMD).}
    \label{fig: diversity parameter}
\end{figure}

\section{EXPERIMENTS}
In this section, we evaluate the performance of our framework in both offline and online evaluation. Here we have used diversity and ranking metrics to evaluate it quantitatively. We have also demonstrated the effectiveness of our framework through the A/B test.
\subsection{Offline Results}
In an industrial setup, off-policy evaluation is challenging. Existing approaches require the data to be randomly logged or the current policy in production to be similar to the test policy\cite{gilotte2018offline,guo2020deep}. However, such assumptions rarely hold true in practical scenarios, as the data received in an industrial setup tends to be purely exploitative and generated by statically positioned widgets for all the customers. We evaluate our algorithm using the traditional offline method where we combine the user and widget contexts and fed it into our framework which generates a ranked order of widgets. We have taken the user widget interaction data for a couple of days to train the model.

\subsubsection{Metrics}
We have used Normalized discounted cumulative gain (nDCG), Intra-list average distance (ILAD)\cite{zhang2008avoiding}, and Intra-list minimal distance (ILMD) as performance metrics. They are defined as:

The nDCG formula is given by:
\[ \text{nDCG} = \frac{\text{DCG}}{\text{IDCG}}, \]

where DCG (Discounted Cumulative Gain) is defined as:
\[ \text{DCG} = \sum_{i=1}^{n} \frac{2^{rel_i} - 1}{\log_2(i+1)}, \]

and IDCG (Ideal Discounted Cumulative Gain) is the DCG score obtained with the ideal ranking of the items.

The Diversity metrics are defined as :
 
\[ \text{ILAD} = \underset{c \epsilon C}{mean}  \underset{i,j \epsilon R_c, i \neq j }{mean} (1-S_{i,j}) \]

\[ \text{ILMD} = \underset{c \epsilon C}{mean}  \underset{i,j \epsilon R_c, i \neq j }{min} (1-S_{i,j}) \]

As we can see from figure \ref{fig: diversity parameter} we captured the impact of varying the trade-off parameter $\theta \epsilon [0,1]$ of  DPP \cite{chen2018fast} w.r.t ILAD and ILMD. Both metrics demonstrate an almost monotonic decreasing behavior with respect to $\theta$, while nDCG exhibits an increasing pattern. Here the higher value of $\theta$ implies more relevance. We identify a promising range, $0.6 \leq \theta \leq 0.8$, where a desirable trade-off between relevance and diversity can be achieved in our widget ranking case.

\begin{figure}[htb]
    \centering
    \includegraphics[scale=0.4]{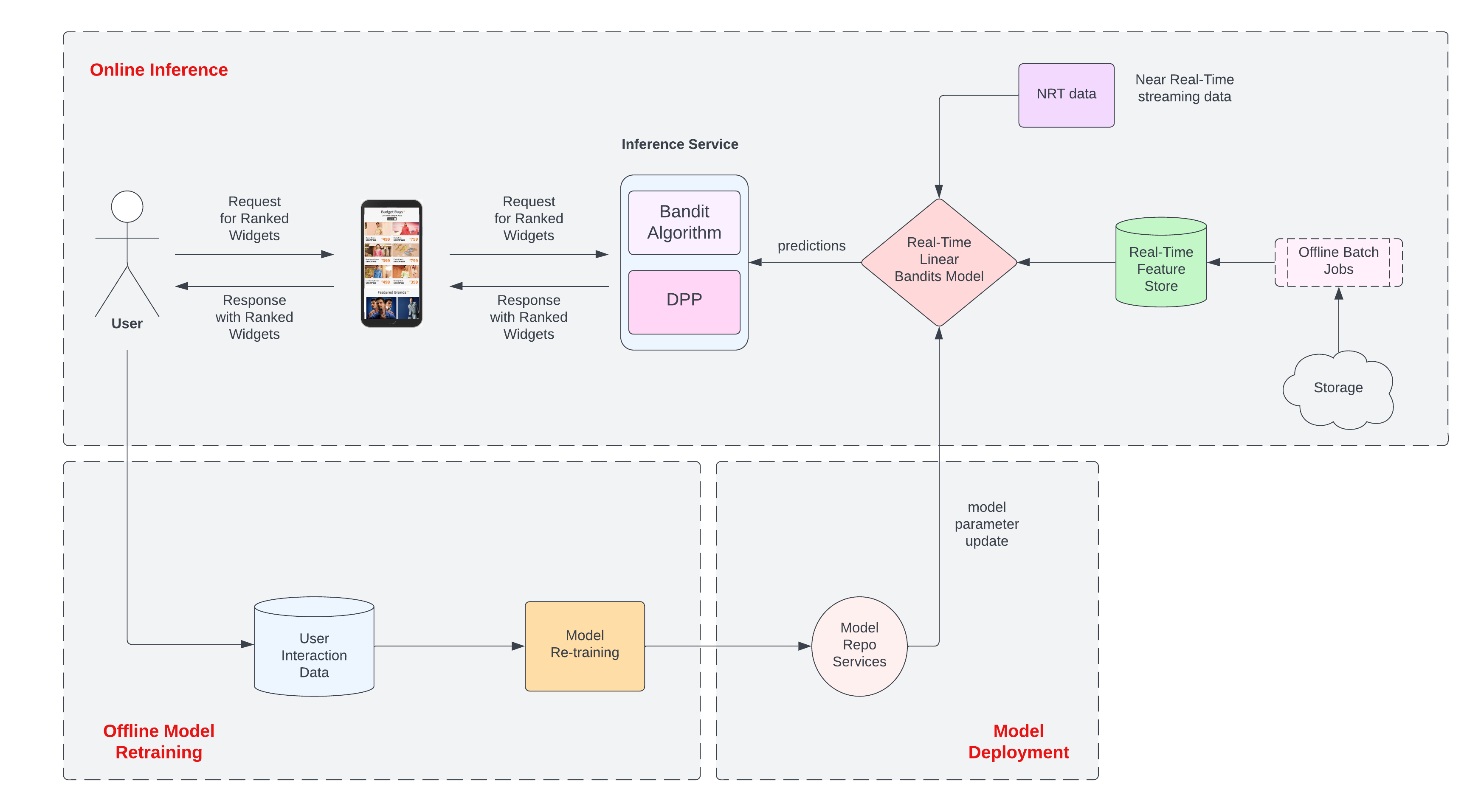}
    \caption{Production system architecture. Our production system consists of (1) online inference that coordinates the frontend request and generates ranked widgets by executing Inference Service, (2) offline model retraining in batch mode after every time 't', and (3) model versioning and deployment}
    \label{Production system architecture}
\end{figure}

\subsection{Online Results}
We warm-started the model using the last 7 days of data before launching the A/B test. We evaluated the XGBoost model alongside linUCB in a production environment. Our results demonstrated an improvement of $\boldsymbol{0.24\%}$ in Home page clicks per user over the XGBoost model with a high statistical significance which is a decent gain given our scale. All other guardrail metrics such as RPU (Revenue per user), user conversion, etc. also exhibited positive outcomes. Figure \ref{Production system architecture} shows the overall production system architecture. 


\section{CONCLUSION AND DISCUSSION}
In this paper, we modeled the personalized widgets ranking as a contextual multi-armed bandit problem focusing on improving the overall shopping experience of the customers. Therein, we addressed the challenges faced in ranking widgets, we highlighted the benefits of our system, notably modeling the widgets with features so as to cater to different types of widgets in a single ranking purview and integrating it with DPP to achieve diversity in ranking. We have shown offline results of our framework where we compared LinUCB and LinTS. This framework can easily be extended to any bandit algorithm and it can be served not just on the homepage but to any other store pages.
Future work can include a multi-objective optimization or defining a reward in such a way that it represents an aggregate of different objectives, we believe this can be a good research direction.

\bibliographystyle{plain}
\bibliography{vwrbib}

\end{document}